\documentclass{ws-ijmpa}

\begin{document}

\markboth{Wei Liang Qian and Ru-Keng Su}
{A quark meson coupling model with density- and temperature- dependent quark masses}

%
\catchline{}{}{}{}{}
%

\title{A quark meson coupling model with density- and temperature- dependent quark masses}

\author{Wei Liang Qian\footnote{wlqian@fudan.edu.cn}}

\address{Department of Physics, Fudan University, Shanghai 200433, P.R.China}

\author{Ru-Keng Su\footnote{rksu@fudan.ac.cn}}

\address{Department of Physics, Fudan University, Shanghai 200433, P.R.China}

\maketitle

\pub{Received (Day Month Year)}{Revised (Day Month Year)}

\begin{abstract}
Based on the quark mass density- and temperature- dependent model
we suggest a model for nuclear matter where the meson field is
introduced to be directly coupled to the quarks.
The dynamic formation of the nucleon bag,
the saturation properties of nuclear matter as well as
equation of state for this model are studies.

\end{abstract}

Relativistic calculation of infinite nuclear matter as well as
neutron matter plays an important role in nuclear physics.
The quantum hadrodynamics(QHD) models\cite{sw86}
based on baryon and meson degress of freedom
form a reasonable starting point for the study of bulk
as well as single-particle properties of nuclei.
However, quantum chromodynamics(QCD)
is believed to be essential in the study of nuclear phenomena
from the quark structure of nucleon. It is also popular
to describe nucleon, hyperon or strangelet in terms of bag models.

The quark mass density- and temperature- dependent
model(QMDTD) is an extension of the original quark mass density-
dependent model(QMDD)\cite{frw81}.
While the latter can almost reproduce the properties of quark matter
obtained by the MIT bag model\cite{bl95},
but it meets many difficulties when we extend this model to finite temperature\cite{zs01}.
Based on Frieberg-Lee soliton bag models\cite{lee81},
it is recognized that the bag constant $B$ decreases with increasing temperature.
It becomes zero and the nontopological solition
solution dissappears at the phase transition point $T_c$.
This dynamic deconfinement mechanism is incorporated in the QMDTD model
by introducing an ansatz $B = B(T) = B_0[1-a(T/T_c)+b(T/T_c)^2]$,
where $B_0$ is the vacuum energy density inside the bag at
zero temperature , $T_c =170 MeV$ is the critical temperature of
quark deconfinement phase transition, and $a$, $b$ are two adjust parameters.
In our previous papers\cite{zs01},
we have investigated the properties of strange matter and strangelets
by using the QMDTD model.

Since the quarks in QMDTD model do not interact with each other,
one can not directly employ this model to investigate the bulk properties of
hadronic system.
Many years ago, the quark-meson coupling(QMC)\cite{g88}
model was proposed to describe nucleon and nuclear matter.
In this model, baryon matter is consisting of non-overlapping MIT bags
bound by the self-consistent exchange of mesons in the mean field approximation.
The quarks are confined in the nucleon bag by the boundary condition.
Inspiring by the QMC model, one can also incorporate the quark-meson
coupling mechanism in the QMDTD model, and employ this model to
investigate the nuclear matter.

The present work evolves from an attempt to extend the QMDTD model
to describe infinite nuclear matter at zero temperature by introducing the
explicit quark meson coupling in the same way as that of the QMC model.
The major difference between the present treatment and the QMC model is that
our model is based on Fridberg-Lee non-topological soliton model and
QMC model is based on the MIT bag model.

In the QMDTD model, the masses of $u$, $d$ quarks
(and the corresponding anti-quarks) are given by \cite{zs01}
\begin{eqnarray}
m_q &=&{\frac{B(T)}{3n_B}}
\label{su1}
\end{eqnarray}
where $n_B$ is the baryon number density, $B(T)$ is the vacuum energy
density.

The quark meson coupling is introduced in the similar way
resembling that of QMC model\cite{g88}.
Let $\bar{\sigma}$ and $\bar{\omega}$ be the mean field values for the
scalar field $\sigma$ and time component of vector field $\omega$, respectively.
The effecitve quark mass $m_q^*$ is defined as
\begin{equation}
m_q^* = m_q - V_{\sigma} = m_q - g_{\sigma}^q\bar{\sigma}
\end{equation}
The energy of the nucleon bag is
\begin{equation}
E_{Bag} = \sum_q E_q = -\sum _q V \int_0^{k_F^q} {\frac{dN_q(k)}{dk}} \varepsilon_q(k) dk \ \ \ \ (q = u, d)
\end{equation}
where $\varepsilon _q(k)=\sqrt{m_q^{*2}+k^2}$ is the single
particle energy, $V_{\sigma} = g_{\sigma}^q\bar{\sigma}$ and
$V_{\omega} = g_{\omega}^q\bar{\omega}$ with the quark-meson coupling
constants, $g_{\sigma}^q$ and $g_{\omega}^q$, $\frac{dN_q(k)}{dk}$ is the density of states\cite{zq01}
for various flavor quarks and $k_F^q$ is the Fermi momentum for quarks.
The stability condition for the bag radius $R$ thus reads
\begin{equation}
\frac{\delta E_{Bag}}{\delta R}=0.  \label{13}
\end{equation}

The total energy per nucleon at the nuclear density $\rho_B$ is given by
\begin{equation}
E_{tot} = \sum _i \frac{2}{\rho_B(2\pi)^3} \int_0^{k_F} d^{3}k\sqrt{M_i^{*2}+k^2}
+ \frac{g_\omega^2}{2m_\omega^2}\rho_B
+ \frac{m_\sigma^2}{2\rho_B}\bar{\sigma}^2
\end{equation}
$m_\sigma$ and $m_\omega$ are the mass for $\sigma$ and $\omega$ mesons
respectively,
the effective mass of nucleon is defined as the bag energy $E_{bag}$,
$n_i$ are the baryon densities for proton and neutron,
$\rho_B$ is the total baryon density and $k_F$ is the Fermi momentum for the nucleons.
The $\omega$ mean field is determined by
\begin{equation}
\bar{\omega} = \frac{V_\omega}{g_\omega^q} = \frac {g_\omega \rho_B}{m_\omega^2}
\end{equation}
where we have defined that the couplings between nucleon and mesons are
$g_\sigma = 3g_\sigma^q$ and $g_\omega = 3g_\omega^q$. The scalar mean
field $\bar{\sigma}$ is determined by self-consistency condition\cite{g88}:
\begin{equation}
\bar{\sigma} = -\frac{2}{m_\sigma^2(2\pi)^3} \sum_i \int_0^{k_F} d^{3}k
\frac{M_i^*}{\sqrt{M_i^{*2}+k^2}}
\left( \frac{\partial{M_i^*}}{\partial{\sigma}} \right)_{R}
\end{equation}

Now we are in a position to present numerical results.
At zero temperature,
the result is irrelevant to the values of $a$ and $b$.\cite{zs01}
We take $B_0 = 173 MeVfm^{-3}$ to fix the mass of nucleon to be $939MeV$.
In Fig.1, the saturation curve for
symmetric nuclear matter is depicted,
where we have choosen the values of $g_\sigma^q$ and $g_\omega^q$ to reproduce
the binding energy $E = -16 MeV$ at $\rho_B = 0.15 fm^{-3}$ for symmetric nuclear
matter at saturation point when $T = 0 K$, and find
\begin{equation}
\frac{g_\sigma^2}{4\pi} = 14.4,\ \ \ \frac{g_\omega^2}{4\pi} = 3.39
\end{equation}
The corresponding equation of state(EOS) for symmetric nuclear matter
is given in Fig.2. 
The coupling constants between quark and mesons and some calculated quantities
such as the effective nuclear mass($M^*$), compressibility($K$), bag radius($R$)
and its relative modification (with respect to its value at zero density) of nuclear
matter at saturation density are listed in Table 1.
To make a comparison, we also enumerate the results obtained
by the QMC model in Table 1.
It is found that these quantities of nuclear matter
in our present model differ not very much from those of the QMC model.

In summary, we have proposed a possible nuclear matter saturation mechanism
based on the QMDTD model.
We follow the QMC model to introduce the quark meson couping under mean field
approximation which provide a reasonable description for nuclear matter.

\begin{figure}
\centerline{\psfig{file=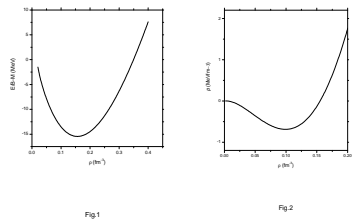,width=5cm}}
\vspace*{8pt}
\caption{Fig.1 Energy per baryon density versus nuclear density
in symmetric nuclear matter at $T = 0 MeV$.
Fig.2 Pressure as a function of baryon density(EOS) for symmetric nuclear
matter at $T = 0 MeV$}
\end{figure}

\begin{table}[h]
\tbl{Comparison of calculated quantities in QMDTD and QMC models.}
{\begin{tabular}{@{}ccccccc@{}} \toprule
      & $g_\sigma^2/4\pi$ & $g_\omega^2/4\pi$ & $M^*$ & $K$ & $R_0$ & $\frac{\delta R}{R_0}$ \\ \colrule
QMC   & 22.0 & 1.14 & 851 & 200 & N/A  & -0.02 \\
QMDTD & 14.4 & 3.39 & 802 & 229 & 0.86 &  0.08 \\
\botrule
\end{tabular}}
\end{table}

\section*{Acknowledgments}

This work is supported in part by NNSF of China under No.10375013, 10347107, 10405008, 10247001,
10235030, National Basic Research Program of China 2003CB716300,
and the Foundation of Education Ministry of China 20030246005.

\end{document}